\documentstyle[preprint,aps,epsf]{revtex}
\newif\iftightenlines\tightenlinesfalse
\tightenlines\tightenlinestrue
\begin{document}
%
\def\pT{p_T^{\phantom{7}}}
\def\MW{M_W^{\phantom{7}}}
\def\ET{E_T^{\phantom{7}}}
\def\bh{\bar h}
\def\lm{\,{\rm lm}}
\def\tG{\tilde G}
\def\lo{\lambda_1}                                              
\def\lt{\lambda_2}
\def\ETC{E_T^c}
\def\pslt{p\llap/_T}
\def\eslt{E\llap/_T}
\def\etmiss{E\llap/_T}
\def\eslt{E\llap/_T}
\def\to{\rightarrow}
\def\Re{{\cal R \mskip-4mu \lower.1ex \hbox{\it e}}\,}
\def\Im{{\cal I \mskip-5mu \lower.1ex \hbox{\it m}}\,}
\def\SU{SU(2)$\times$U(1)$_Y$}
\def\te{\tilde e}
\def \tlam{\tilde{\lambda}}
\def\tl{\tilde l}
\def\tb{\tilde b}
\def\tst{\tilde t}
\def\tt{\tilde t}
\def\ttau{\tilde \tau}
\def\tmu{\tilde \mu}
\def\tg{\tilde g}
\def\tga{\tilde \gamma}
\def\tnu{\tilde\nu}
\def\tell{\tilde\ell}
\def\tq{\tilde q}
\def\tw{\widetilde W}
\def\tz{\widetilde Z}
\def\cmsec{{\rm cm^{-2}s^{-1}}}
\def\fb{{\rm fb}}
\def\sgn{\mathop{\rm sgn}}
\def\mhf{m_{\frac{1}{2}}}

\hyphenation{mssm}
\def\ds{\displaystyle}
\def\ts{${\strut\atop\strut}$}
%
%
\preprint{\vbox{\baselineskip=14pt%
   \rightline{FSU-HEP-980509}\break 
   \rightline{UH-511-902-98}\break
   \rightline{BNL-HET-98/23}
}}
\title{LHC Reach For Gauge Mediated  
Supersymmetry Breaking Models Via Prompt Photon Channels}
\author{Howard Baer$^1$, Pedro G. Mercadante$^2$, Frank Paige$^3$,
Xerxes Tata$^{2}$ and Yili Wang$^2$}
\address{
$^1$Department of Physics,
Florida State University,
Tallahassee, FL 32306, USA
}
\address{
$^2$Department of Physics and Astronomy,
University of Hawaii,
Honolulu, HI 96822, USA
}

\address{$^3$Physics Department,
Brookhaven National Laboratory, Upton, NY 11973, USA
}%
\date{\today}
\maketitle
\begin{abstract}
We evaluate the supersymmetry reach of the Large Hadron Collider within
the gauge-mediated low energy supersymmetry breaking framework, assuming
that a neutralino is the second lightest sparticle and that it decays
promptly into a gravitino which escapes detection.  We find that the maximum
reach is obtained via a search for inclusive $\gamma\gamma+\eslt$ events
coming dominantly from chargino and neutralino production.  Assuming an
integrated luminosity of 10~$fb^{-1}$, we find that LHC experiments will
be able to probe values of the model parameter $\Lambda \alt 400$~TeV,
corresponding to $m_{\tg} \leq 2.8$~TeV. A measure of the  model
parameter $\Lambda$ may be possible from the photon
$p_T$ spectrum.

\end{abstract}

\medskip


Most phenomenological analyses~\cite{REV} of supersymmetry (SUSY) have
been carried out within the framework of the mSUGRA model~\cite{NILLES},
where gravitational interactions induce effects of SUSY breaking
(assumed to occur in a ``hidden sector'') into the ``observable sector''
of ordinary particles and their superpartners. In this case, the
particle-sparticle mass gap in the observable sector is $M_{obs} \sim
\langle F \rangle/M_{Planck}$, where $\sqrt{\langle F \rangle}$ is the scale at
which SUSY is broken. To ensure that SUSY can stabilize the elementary
scalar electroweak symmetry breaking sector, we require $M_{obs} \sim
250$~GeV, and so choose $\langle F \rangle \sim$ (10$^{11}$~GeV)$^2$.
The gravitino, which acquires a mass $\sim M_{obs}$, then decouples from
the observable sector, and is unimportant for particle physics
phenomenology.

Recently, however, there has been a resurgence of interest
\cite{DINE,THOM,AMBROS,BABU,GMLESB,BAGGER,NANDI,GUN,FENGMOR,GMREV} in the
phenomenology of a class of models where usual gauge interactions, not
gravity, serve as the messengers of SUSY breaking. Here, it is assumed
that there is a new set of particles that directly couple to the SUSY
breaking sector, and which also have non-vanishing SM interactions.  The
idea then is that these new particles, which first feel the effects of
SUSY breaking, convey these down to the observable sector via their
gauge interactions.  This framework differs from the mSUGRA framework in
two important ways.  First, because SUSY breaking is fed down to the
observable sector by gauge interactions, the mass difference between
ordinary particles and their superpartners depends on their gauge
interactions -- gluinos are thus heavier than $SU(2)$ gauginos which are
heavier than binos, and $m_{\tq} > m_{\tell_L} \sim m_{\tnu} >
m_{\tell_R}$. Second, since gravitational interactions are not the
messengers of SUSY breaking, $M_{obs}$ is not suppressed by $M_{Planck}$
as in mSUGRA. Instead, we now have sparticle masses $\sim \alpha \langle
F_M \rangle /M$, where M is the mass scale of the messenger particles
and $\langle F_M \rangle$ the induced SUSY breaking scale in the
messenger sector, with $\alpha$ the relevant gauge coupling of the
sparticle. The important thing is that it is possible to get weak scale
sparticle masses with $\langle F_M \rangle \sim 100$~TeV if $M$ is
100-1000~TeV. Indeed it could well be that the intrinsic SUSY breaking
scale $\langle F \rangle$ is of the same magnitude, and supersymmetry
breaking is determined by 100~TeV scale physics. If this is the case,
then the gravitino mass $m_{\tG} = {\langle F
\rangle}/{(\sqrt{3}M_P)}$ (here $M_P = M_{Planck}/\sqrt{8\pi}$) could be
${\cal {O}}(1 \ eV)$. This is important because (the longitudinal
components of) such superlight gravitinos have phenomenologically
relevant couplings to ordinary particles~\cite{FAYET}.

In the specific model that we adopt for our analysis, the messenger
sector comprises of $n_5$ sets of ``quark'' and ``lepton'' superfields
in a 5+$\bar{5}$ representation of $SU(5)$, with superpotential
couplings to a gauge singlet superfield ($\hat S$) which acquires vacuum
expectation values for both its scalar ($S$) and auxiliary ($F_M$)
components. The former results in masses for messenger quark and lepton
superfields, whereas the latter is the result of the feed down of SUSY
breaking from the hidden sector. Assuming that $\langle F_M \rangle \ll
\langle S \rangle ^2$, the gauginos acquire a mass
\begin{equation}
m_{\tlam_i} = \frac{\alpha_i}{4\pi}n_5\Lambda,
\label{eq:gaugino}
\end{equation}
whereas scalar components of chiral superfields get a mass,
\begin{equation}
m_{scalar}^2=2n_5\Lambda^2 \left [ C_3(\frac{\alpha_3}{4\pi})^2 +
C_2(\frac{\alpha_2}{4\pi})^2
+\frac{3}{5}(\frac{Y}{2})^2(\frac{\alpha_1}{4\pi})^2\right ],
\label{eq:scalar}
\end{equation}
with $\Lambda=\langle F_M \rangle /\langle S \rangle$, and $\alpha_1$
given in terms of the usual hypercharge coupling $g'$ by
$\alpha_1=\frac{5}{3}\frac{g'^2}{4\pi}$. Finally, $C_3=\frac{4}{3}$ for
colour triplets and zero for colour singlets while $C_2=\frac{3}{4}$
for weak doublets and zero for weak singlets. These relations, which
are independent of the messenger sector superpotential couplings $\lambda_i$
get corrections of
$\sim {\langle F_M\rangle}/({\lambda_i \langle S^2 \rangle})$ which
are ignored in the subsequent analysis. 
The gravitino mass is given by,
\begin{equation}
m_{\tG}=\frac{\langle F \rangle}{\sqrt{3}M_P} = 
\frac{\langle F \rangle}{\lambda \langle F_M
\rangle}\frac{M\Lambda}{\sqrt{3}{M_P}}
\equiv C_{Grav}\frac{M\Lambda}{\sqrt{3}{M_P}}.
\end{equation}
The parameter $C_{Grav}$ that we have introduced is $\geq 1$ since the
SUSY breaking scale in the messenger sector cannot be larger than the
intrinsic SUSY breaking scale in the hidden sector, and the
superpotential coupling $\lambda \leq 1$.  SUSY breaking $A$-parameters
and the $B$-parameter are induced only at two loops so that it is
reasonable to suppose that these are small.  The supersymmetric $\mu$
parameter is not determined by how SUSY breaking is mediated but is
fixed (up to a sign) by radiative symmetry breaking as in the mSUGRA
framework.\footnote{It is customary to eliminate $\tan\beta$ in favour
of $B_0$, the $B$ parameter at the scale $M$. Since $B_0$ is small in
this framework, this constrains $\tan\beta$ to be 20-30, the precise
value depending on the choice of $\Lambda$. As in Ref.\cite{GMLESB}, we
do not impose this constraint since we expect that it will be altered by
new interactions introduced to generate $\mu$ dynamically.}  A complete theory
that includes the dynamics of SUSY breaking will presumably yield a
value of $\mu$ consistent with this.  Eq. (\ref{eq:gaugino}) and
Eq. (\ref{eq:scalar}) should be regarded as boundary conditions for the
gaugino and scalar masses valid at the messenger scale $M=\lambda
\langle S \rangle$, so that these parameters need to be evolved down to
the weak scale relevant for phenomenological analysis. Our framework is
thus completely parametrized by\footnote{It is possible to
construct~\cite{THOM} 
more complicated models which lead to different boundary conditions for
scalar and gaugino masses from Eq.~(\ref{eq:gaugino}) and
Eq.~(\ref{eq:scalar}). We do not consider such scenarios in our analysis.}
\begin{displaymath}
(\Lambda, M, n_5, \tan\beta,\sgn\mu, C_{Grav})
\end{displaymath}
which together with Standard Model parameters completely determine the
masses and couplings of all particles in the observable sector. The
messenger mass scale $M$ can range from $\Lambda$ to $M_{GUT}$ or
$M_P$. We focus our attention on models where $M \sim 1$~PeV, so that the
SUSY breaking scale $\langle F_M \rangle$ (and possibly $\langle
F \rangle$) may be relatively low.

In order to facilitate simulation of models with gauge mediated
supersymmetry breaking (GMSB) we have recently included in
ISAJET v7.37 \cite{ISAJET} a `GMSB option' that allows one to use
the GMSB parameter set introduced above as an input. ISAJET then
evolves the gaugino and scalar masses as given by Eq.~(\ref{eq:gaugino})
and Eq.~(\ref{eq:scalar}) from the scale $M$ down to the scale relevant
for phenomenology, and computes the ``MSSM parameters'' that are then
used in the
evaluation of sparticle cross sections and decay patterns. The decays
of neutralinos into gravitinos, $\tz_i \to \tG \gamma$, $\tz_i \to \tG
Z$ and $\tz_i \to \tG h,H,A$ as well as (approximately) the Dalitz decay
$\tz_i \to e^+e^-\tG$ are included in ISAJET. The decays $\tell_R \to
\ell\tG$ and $\ttau_1 \to \tau\tG$, as well as the three body decays~\cite{AMK}
(mediated by a virtual neutralino) $\tell_R \to \ttau_1 \bar{\tau}\ell$
and $\tell_R \to \bar{\ttau_1}\tau \ell$ have also been included. Widths
for corresponding three body decays mediated by virtual chargino
exchange are suppressed by the lepton Yukawa coupling, and have not yet been
included. These can be significant only when $m_{\tmu_R}-m_{\ttau_1}
\alt m_{\tau}$ so that the neutralino-mediated three body decays of
$\tmu_R$ are kinematically very suppressed or forbidden.

Within the GMSB framework, the parameter $\Lambda$ sets the scale of
sparticle masses. The dependence on the messenger scale~($M$), which
specifies where the boundary conditions (\ref{eq:gaugino}) and
(\ref{eq:scalar}) hold is presumably logarthmic. To help the reader
link the model parameters with the more familiar sparticle masses, we
show in Fig.~1 contours of $m_{\tg}$ in the $\Lambda-\tan\beta$ plane
for ({\it a})~$n_5=1$, and ({\it b})~$n_5=2$. We fix $m_t=175$~GeV, $M
= 1$~PeV and take $\mu>0$. We have checked that the spectrum is only
weakly sensitive to $M$ for a range of $M$ between $\Lambda$ and
1~PeV, and also that it is insensitive to the sign of $\mu$ except for
very low values of $\tan\beta$. The $SU(2)$ and $U(1)$ gaugino masses
are roughly given by $\frac{1}{3}m_{\tg}$ and $\frac{1}{6}m_{\tg}$,
respectively. The scales of the sfermion masses can roughly be read
off from Eq.~(\ref{eq:scalar}).  In Region I below the diagonal solid
line~\footnote{There are finite one loop corrections~\cite{REF} which
alter the relationship between the fermion mass and the corresponding
Yukawa coupling that have not been included in ISAJET.  These
corrections, which can be important for large $\tan\beta$, would shift
the exact location of this line. Thus, for model parameter values
close to this line, these corrections could radically alter the
phenomenology if the NLSP shifts from being a neutralino to being a
stau.  While these corrections are crucial to include in any study relating
SUSY phenomena to the underlying parameters (that may be determined
by physics at yet higher energy scales)
of the theory, 
this issue is not critical to our main point which is a study
of the LHC reach for the case where $\tz_1$ is the NLSP.}  
in frame~({\it a}), $m_{\tz_1} < m_{\ttau_1}$, so that
$\tz_1 \to \tG\gamma$ (this dominates the decays to $Z$ or Higgs
bosons), so that isolated photons are a hallmark of SUSY within this
framework. In Region II, sparticles cascade decay to $\ttau_1$ (except
immediately above this line where $m_{\tz_1}-m_{\ttau_1}< m_{\tau}$),
which then decays to $\tau \tG$, so that for this portion of the
parameter space SUSY events contain $\tau$ leptons in addition to
other leptons and jets~\cite{NANDI}.  In frame ({\it b}), $\tz_1$ is
once again the next to lightest SUSY particle (NLSP) in Region I,
while in the other regions $\ttau_1$ is the NLSP (though $m_{\te_R}$
and $m_{\tmu_R}$ may be very close to $m_{\ttau_1}$).  In Region II,
$m_{\tell_R} > m_{\tz_1} > m_{\ttau_1}$, while in Regions III and IV,
$\tz_1 \to \tell_R \ell$ is kinematically possible. In Region III,
$\tell_R \to \ell \tG$ because the decay $\tell_R \to \tau \ttau_1
\ell$ is kinematically inaccessible, while in Region IV, both these
decays are possible and compete. The phenomenology will clearly be
sensitive to just where we are in the parameter space.

The grey region on top of each frame is excluded because proper electroweak
symmetry breaking is not obtained: the wedge in the upper left
corner is excluded because $m_{\ttau_R}^2 <0$ while in the band on the
top, $m_A^2 < 0$. The various hatched regions are excluded by
the non-observation of sparticles or a Higgs boson at LEP 2. In the
region with vertical shading, $m_{\ttau_1}< 60$~GeV (in the
corresponding portion of Region I, this limit may be stronger as
the final state also contains photons); in that
shaded horizontally, $m_{\tw_1}<88$~GeV; in the small diagonally
shaded region at low $\tan\beta$, $m_h < 89$~GeV (the current lower limit on
the SM Higgs boson mass). While these limits~\cite{EXPREV} are essentially 
those obtained in
the mSUGRA model, the strongest constraint comes from
the non-observation of acollinear photon pairs from $e^+e^- \to \tz_1\tz_1 \to
\gamma\gamma\tG\tG$, and is special to this framework. 
The ALEPH collaboration \cite{ALEPH} has excluded
$m_{\tz_1} < 71$~GeV assuming that $\tz_1 \approx \tilde{B}$ and
$m_{\te_R} = 1.5 m_{\tz_1}$, which is a good approximation for
$n_5=1$. This exclusion is shown as  the diagonally shaded vertical
strip in the figures.\footnote{For $n_5=2$ in Fig.~1{\it b}, although we
have shown the same limit of 71~GeV, the reader should be aware that
this is overly conservative because, for any given value of
$m_{\tz_1}$, $m_{\te_R}$ will be lighter than for $n_5=1$, and the
$\tz_1\tz_1$ cross section will be correspondingly larger.} The D0
experiment at the Tevatron~\cite{DZERO}, from a non-observation of
di-photon events with large $\eslt$ have inferred a lower bound $\sim
m_{\tw_1} > 150$~GeV, which essentially excludes this same region. 

We have already noted that if model parameters happen to be in Region I
of Fig.~1, in addition to multijet and multilepton events expected in
the mSUGRA framework, SUSY events would contain, in addition, hard
isolated photons, provided that the SUSY breaking scale $\langle F
\rangle$ is not much larger than $\langle F_M \rangle$ for the
messenger sector, {\it i.e.}, that $C_{grav}$ is not too large.
In Region~II (and III and IV in frame {\it b}), SUSY
events would contain additional $\tau$ (and possibly $e$ or $\mu$)
instead of photons.  The main purpose of this study is to assess the
SUSY reach of the Large Hadron Collider within the GMSB framework,
assuming that the parameters are in Region I. For this reason, we will
mainly focus on the case $n_5=1$ in the rest of the analysis and defer
the study of the phenomenology where sleptons are lighter than $\tz_1$
to a subsequent analysis.

We begin by showing the production cross sections as a function of the
parameter $\Lambda$ for a variety of
sparticle production reactions in Fig.~2. We have fixed $M = 1$~PeV,
$n_5=1$, $\mu>0$ which we take to be our default values, and taken
$\tan\beta=2$. The cross sections are independent of $C_{Grav}$. We use
CTEQ3L structure functions~\cite{CTEQ} for our analysis. We
see that squark and gluino production dominates for small values of
$\Lambda \alt 175$~TeV, beyond which electroweak production of (lighter)
charginos and neutralinos is the dominant source of sparticles. The
production of charginos or neutralinos in association with gluinos or squarks
is always sub-dominant. The dotted curve, which includes cross sections for
$\tell\tell, \tell\tnu$ and $\tnu\tnu$ production, shows that
slepton pair production is an order of magnitude
smaller than chargino-neutralino production. Except when $\Lambda \simeq
50-100$~TeV where $\tnu \tell_L$ production is also important, slepton 
production is dominated by $\tell_R\tell_R$ production ($\tell =e,\mu$ in the
figure), a reflection of the fact that $2m_{\tell_R} \simeq m_{\tell_L} \sim 
m_{\tnu}$. It is interesting to observe that $\tell_R\tell_R$ production
which results in the essentially SM background-free
$\ell\ell\gamma\gamma+ \eslt$ events
may be observable (if they can be separated from other SUSY sources of
dileptons) for $\Lambda$ values as large as 225~TeV,
corresponding to $m_{\tell_R} = 400$~GeV. 

For detector simulation at the LHC, we use the toy calorimeter
simulation package ISAPLT. We simulate calorimetry covering $-5<\eta
<5$ with cell size $\Delta\eta\times\Delta\phi =0.05\times 0.05$. We
take the hadronic energy resolution to be $50\% /\sqrt{E}\oplus 0.03$
for $|\eta |<3$, where $\oplus$ denotes addition in quadrature, and to
be $100\% /\sqrt{E}\oplus 0.07$ for $3<|\eta |<5$, to model the
effective $p_T$ resolution of the forward calorimeter including the
effects of shower spreading, which is otherwise neglected. We take
electromagnetic resolution to be $10\% /\sqrt{E}\oplus 0.01$.
Although we have included these resolutions, which are typical of
ATLAS\cite{ATLAS} and CMS\cite{CMS}, we have made no attempt to
estimate the effects of
cracks, edges, and other problem regions. Much more detailed detector
simulations are needed to understand the effects of such regions and
of the resulting non-Gaussian tails, particularly on the $\eslt$
resolution.

Jets are found using fixed cones of size $R=\sqrt{\Delta\eta^2
+\Delta\phi^2} =0.7$ using the ISAJET routine GETJET. Clusters with
$E_T>100$ GeV and $|\eta ({\rm jet})|<3$ are labeled as jets.  Muons and
electrons are classified as isolated if they have $p_T>10$ GeV, $|\eta
(\ell )|<2.5$, and the visible activity within a cone of $R =0.3$ about
the lepton direction is less than $E_T({\rm cone})=5$ GeV.  Photons with
$E_T > 10$~GeV and $|\eta|< 2.5$ are considered to be isolated if there
is less than 2.5~GeV of other activity in a cone of $R=0.3$ about the
photon. 

To enhance the SUSY signal over SM background processes, we require that
each event contains at least two entities (an entity is a jet, an
electron, a muon or a photon) with $E_T > 100$~GeV and $\eslt > 100$~GeV
and at least one isolated photon with $E_T > 40$~GeV. The inclusive two photon
sample is required to have a second isolated photon with $E_T >
25$~GeV. As in previous analyses~\cite{LHC1,LHC2}, we find it useful to
use variable cuts for the two hardest entities and $\eslt$: we require
$E_{T1}, E_{T2}, \eslt \geq E_T^c$, where we vary the parameter $E_T^c$.

To estimate SM background we have not done a new simulation, but used
our previous computations of backgrounds for SUSY signals {\it without
photons} to obtain an estimate of the background level for the present
situation. The background cross sections as obtained from Fig.~3 of
Ref. \cite{LHC2} are shown in the first three rows of Table~1 for
several values of $E_T^c$.  In this computation, the basic lepton and
jet identification requirements were identical to the ones used in the
present analysis and $E_T(j1),E_T(j2)$ and $\eslt$ were all required to
be larger than $E_T^c$. Radiation of each hard isolated ({\it i.e.}
large angle) photon is suppressed by a factor $\alpha$ (we neglect
coefficients of ${\cal{O}}(1)$ and factors of $2\pi$), so that we may
estimate the {\it physics} backgrounds for SUSY events with one or two
isolated photons by multiplying the total background in Ref.~\cite{LHC2}
by $\alpha$ and $\alpha^2$, respectively. This background level is shown
in the next two rows of the Table.  What about non-physics backgrounds
from mis-identification of jets or leptons as photons? We estimate this
by assuming that the jet gamma rejection is 1/5000 \cite{REJECT}, and
that SUSY events typically have a jet multiplicity $\sim 10$. The chance
of an electron to fake a photon is much smaller ($\sim 10^{-5}$) so that
this background is negligible.  We may thus obtain a crude estimate of
the non-physics backgrounds to single (two) photon events by reducing
the total background in Ref.~\cite{LHC2} by a factor of 500 ($
\frac{45}{5000^2} \sim 2\times 10^{-6}$). This is shown in the last two
rows of Table~1.  An important conclusion that one draws from the Table
is that while the detector-dependent background to single photon events
may be comparable to the physics background, it is relatively
unimportant for the inclusive two photon SUSY event sample.

The reader may worry that the background sample always contains two jets
with $E_T>E_T^c$, while a portion of the signal may instead contain
leptons or photons instead of jets as the primary entities, the
background to which has not been included in Ref. \cite{LHC2}. While we
have not estimated this, we do not expect it to be a problem since it is
reasonable to suppose that in the SM, events with very hard isolated
leptons and/or photons are much rarer than those with hard jets.

In order to estimate the SUSY reach of the LHC, we have computed the
signal cross sections for the one and two-photon inclusive SUSY samples.
Typically, we found that after acceptance cuts, the two photon inclusive
signal rate is suppressed relative to the corresponding single photon
rate by a factor of $2-3$, depending on sparticle masses. Since two
photon backgrounds are much more strongly suppressed the best reach is
obtained in the $\gamma\gamma+\eslt + X$ channel.  We show LHC cross
sections for such events versus $\Lambda$ after cuts in Fig.~3 for ({\it
a})~$\mu > 0, \tan\beta=2$, ({\it b})~$\mu > 0, \tan\beta = 35$, and
({\it c})~$\mu < 0, \tan\beta=2$. The parameters $n_5$ and $M$ 
are set at our default values. In our computation, we
assume that the decay $\tz_1 \to \gamma\tG$ occurs instantaneously; {\it
i.e.} the photon emerges from the primary vertex. (The typical decay
length is millimeters for $C_{grav}=1$. This is short enough to be
unimportant for photons but is easily measured via $\tz_1\to e^+ e^-\tG$
Dalitz decays.)
Also shown is the
background level, estimated as described above. We show cross sections
for $E_T^c =100$~GeV (solid) and 200~GeV (dashed). The scale on the top
axis is the gluino mass, which is essentially the same for each of the
frames. We see that by choosing the larger value of $E_T^c$, especially
for $\Lambda \agt 200-250$~GeV, the SM background can be greatly
reduced with only modest loss of signal. Indeed, the signal is then only
rate limited, and we obtain a 5 event level, which we take to define the
reach of the LHC, with an integrated luminosity of 10~$fb^{-1}$ if
$\Lambda \alt 400$~TeV (corresponding to $m_{\tg} \alt 2.8$~TeV) in all
three frames. The expected SM background is 0.5 event, so that the
Poisson probablility of fluctuation to give more than five events is
$1.4 \times 10^{-5}$. Of course, one must view this number in
perspective, considering the crudeness of our background estimate. Note
however, that even if the background is larger than our estimate by an order of
magnitude, the typical experiment would exclude the SM at 99\%~CL if
$\Lambda \alt 380$~TeV, or a ``$5\sigma$ discovery'' would be possible
for $\Lambda \alt 300$~TeV.

We remark that we have also checked for several cases along
the boundary of Region I in Fig.~1{\it b} that the SUSY signal is
observable even if $n_5=2$, as long as $\tz_1 \to \tG\gamma$ well inside
the detector. We expect that the signal is observable over this entire region.

Finally, we briefly consider prospects for determining the parameter
$\Lambda$ that sets the sparticle mass scale in our scenario. At first,
it might seem that the mean values of the total scalar energy, the
visible invariant mass, or the $E_T$ of the hardest entities in SUSY
events would scale with $\Lambda$. This is, however, not the case
because the dominant SUSY sources change from $\tg$ and $\tq$ to (much lighter)
charginos and neutralinos as $\Lambda$ increases from below 200~TeV to
above 200~TeV. The mass of $\tz_1$ (which is dominantly a bino), on the
other hand always scales with $\Lambda$, so that it is reasonable to
expect that the $p_T$ of the photons produced via its decays to the
gravitinos would scale similarly. To illustrate this we show in Fig.~4
the $p_T$ distribution of the hard photon in SUSY events where both
photons are detected for several values of $\Lambda$. For definiteness,
we have illustrated this for four values of $\Lambda$ with other
parameters as in Fig.~3{\it a}, and for $E_T^c = 100$~GeV. Assuming that
our background estimate is valid, we see that the distributions show
a clear scaling with $\Lambda$, though for the highest value of
$\Lambda$ an integrated luminosity of 30-50~$fb^{-1}$ would be required
to construct this distribution. The $p_T$ distribution of the second
photon shows the same scaling behaviour.

Although the $p_T(\gamma)$ distributions in Fig.~4 scale nicely with
$\Lambda$, it is not enough to determine $\Lambda$ as $n_5$ is not
known. The most obvious strategy for the determination of $n_5$ is to
obtain the ratio of sfermion and gaugino masses. This may, for example,
be possible by focussing on event chains~\cite{HINCH} in which the
decays $\tz_{2} \to \ell\tell_R \to \ell \bar{\ell}\tz_1 \to \ell
\bar{\ell}\gamma\tG$ (for $n_5=1$) or even $\tz_1 \to \tell_R \ell\to
\ell\bar{\ell}\tG$ (for $n_5=2$) can be isolated. Additional information
may also be available from the overall mass and $E_T$ of the hardest
entities in SUSY events.  The distribution of decay lengths of the NLSP,
which may be possible to determine from the displacement of the
secondary vertex can yield information about the underlying SUSY
breaking scale.  A study of these issues, and also the reach of the LHC
when the NLSP is not the lightest neutralino, is in progress.

{\bf Acknowledgement:} We are grateful to Uri Sarid and Carlos Wagner
for bringing the importance of the corrections in Ref. \cite{REF} to
our attention. P.M. was partially supported by 
Funda\c{c}\~ao de Amparo \`a Pesquisa do Estado de
S\~ao Paulo (FAPESP). 
HB thanks the Davis Institute for High Energy Physics for hospitality
while a portion of this work was completed. 
This research was supported in part by
U.S. Department of Energy Grants DE-FG-03-94ER40833,
DE-FG-02-97ER41022 and DE-AC02-76CH00016.

\newpage				   

\begin{center}
\begin{table}[ ]
\caption[]{SM background cross sections from Fig.~3 in Ref. \cite{LHC2}
as a function of $E_T^c$ defined in the text. The $2\ell$ background
includes contributions from both same sign and opposite sign
leptons. Trilepton background cross sections are much smaller. The next
two rows, which are obtained by multiplying the total background by
$\alpha \sim 10^{-2}$ and $\alpha^2 \sim 10^{-4}$, give our estimate of
physics sources for single photon and two photon inclusive events
passing the SUSY selection cuts. Detector-dependent backgrounds, which
are estimated as discussed in the text, are shown in the last two rows.}
\bigskip
\begin{tabular}{|l|c|c|c|}
$E_T^c$~(GeV) & 100 & 200 & 300 \\ 
\tableline
$\sigma(\eslt)$~($fb$) & $10^4$ & 500 & 50 \\ 
$\sigma(1\ell)$~($fb$) & 1200 & 30 & 4 \\ 
$\sigma(2\ell)$~($fb$) & 700 & 10 & 0.4 \\ \hline
$\alpha\times\sigma_{Tot}$~($fb$) & 100 & 5 & 0.5 \\
$\alpha^2\times\sigma_{Tot}$~($fb$) & 1 & 0.05 & 0.005 \\ \hline
Fake($1\gamma$)~($fb$) & 20 & 1 & 0.1 \\
Fake($2\gamma$)~($fb$) & 0.02 & $10^{-3}$ & $10^{-4}$ \\
\end{tabular}
\end{table}
\end{center}

%

%
%
%


\iftightenlines\else\newpage\fi
\iftightenlines\global\firstfigfalse\fi
\def\dofig#1#2{\iftightenlines\epsfxsize=#1\centerline{\epsfbox{#2}}\bigskip\fi}

\begin{figure}
\dofig{5in}{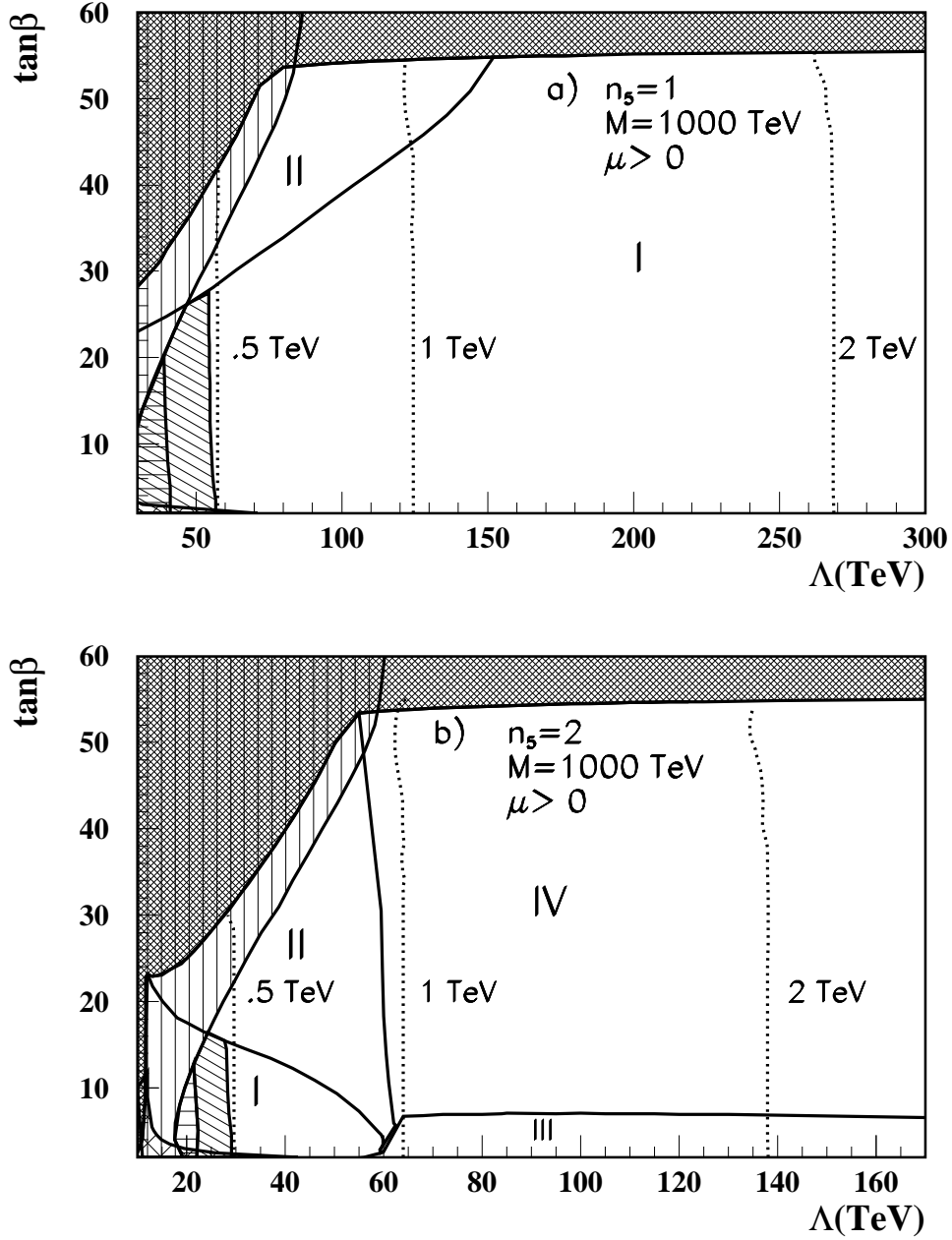}
\caption[]{Various regions in the $\Lambda-\tan\beta$ plane for
$M=1000$~TeV and $\mu> 0$ for {\it a})~$n_5=1$ and {\it b})~$n_5=2$. The
grey region is excluded by theoretical considerations while the various
hatched regions are excluded by experimental constraints discussed in
the text. Also shown by the dotted curves are contours of gluino mass.
In Region I, $\tz_1$ is the NLSP while in the other regions, the NLSP is
a slepton.\label{fig1}}
\end{figure}

\begin{figure}
\dofig{5in}{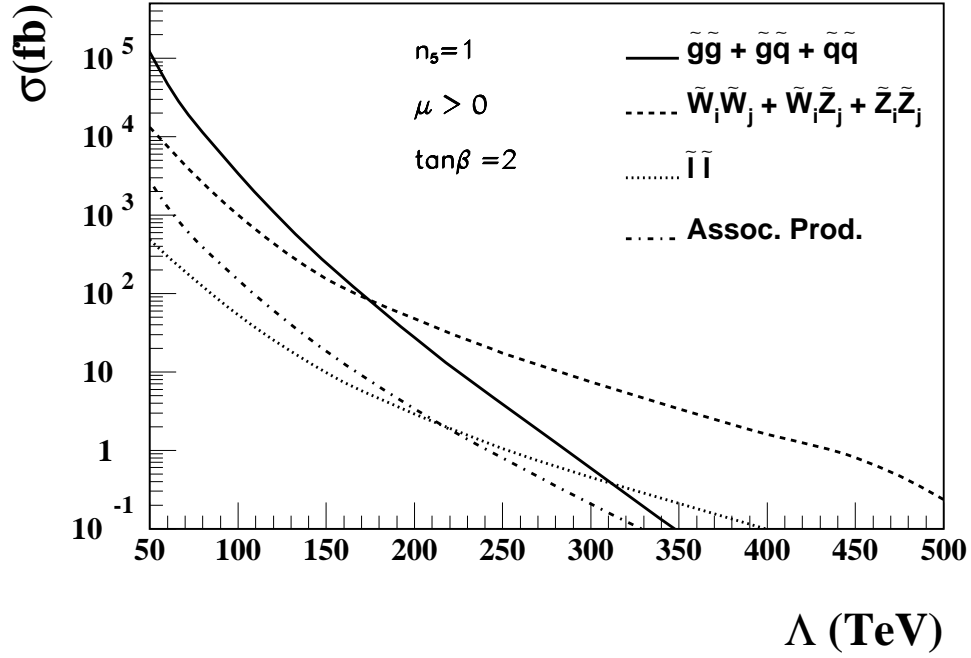}
\caption[]{The production cross sections versus $\Lambda$ for various
sets of SUSY processes at a 14~TeV $pp$ collider within the GMLESB
framework. We show the cross sections for $M=1000$~TeV, with other
parameters as shown in the figure. For sleptons, we sum over the first
two families.\label{fig2}}
\end{figure}

\begin{figure}
\dofig{3.5in}{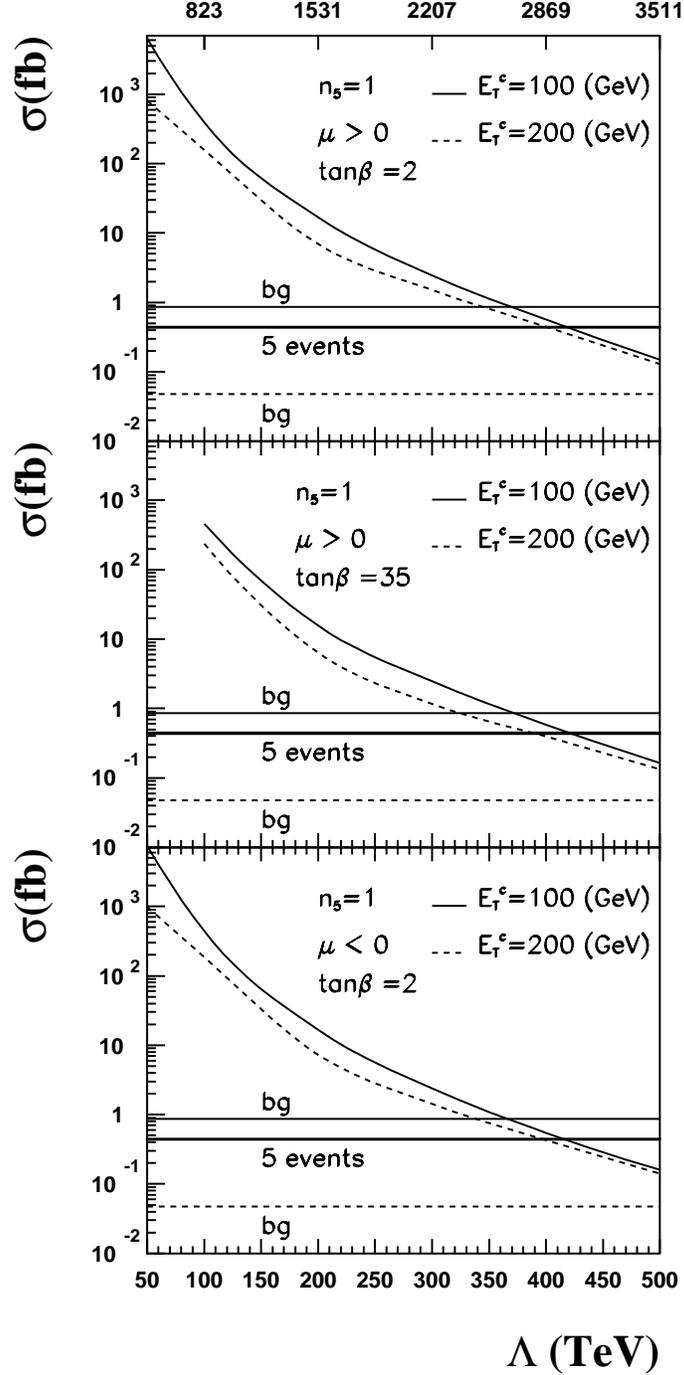}
\caption[]{Signal cross sections for inclusive $\gamma\gamma+\eslt+X$ events at
the LHC after the cuts described in the text for two values of the cut
parameter $E_T^c$. We fix
$M=1000$~TeV. Our estimate of the background level from SM sources is
shown by the lines denoted by bg. The five event level is also shown for
an integrated luminosity of 10~$fb^{-1}$.\label{fig3}}
\end{figure}

\begin{figure}
\dofig{5in}{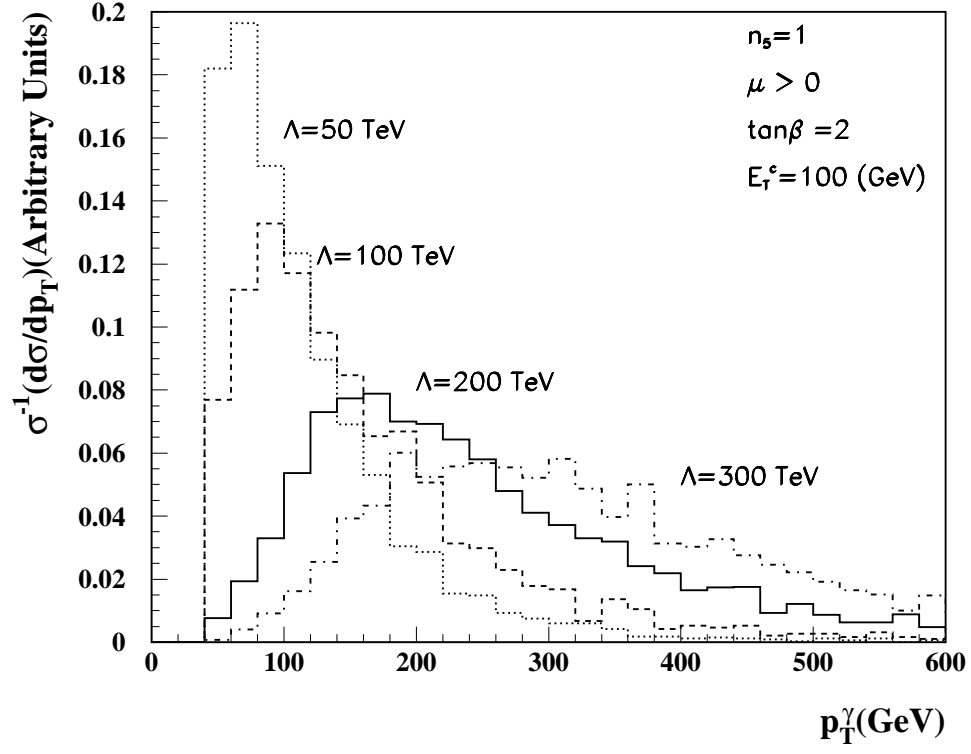}
\caption[]{The transverse momentum distributions for the hard photon in
inclusive $\gamma\gamma+\eslt$ events at the LHC as computed within the
GMSB framework for several values of $\Lambda$ and the cut parameter
$E_T^c=100$~GeV. We choose $M=1000$~TeV.\label{fig4}}
\end{figure}

\vfil

\end{document}